\documentclass[aps,pra,twocolumn,reprint,superscriptaddress,showpacs]{revtex4-1}

\usepackage{graphicx}
\usepackage{bm}
\usepackage{amsmath}
\usepackage{color}
\usepackage[colorlinks=true,pdfborder={0 0 0},citecolor=blue,linkcolor=blue,urlcolor=blue]{hyperref}

\newcommand{\SqrtBy}[2]{$\sqrt{#1}$\kern0.2ex$\times$\kern-0.2ex$\sqrt{#2}$}

\newcommand{\uVrms}{$\mu$V$_{\rm rms}$}
\newcommand{\EF}{E_{\rm F}}
\newcommand{\vF}{v_{\rm F}}
\newcommand{\kF}{k_{\rm F}}
\newcommand{\kB}{k_{\rm B}}
\newcommand{\Tc}{T_{\rm c}}

\newcommand{\Bext}{B_{\rm ext}}

\newcommand{\AffiliationMANA}{International Center for Materials Nanoarchitectonics, National Institute for Materials Science, 1-1 Namiki, Tsukuba 305-0044, Japan}
\newcommand{\AffiliationISSP}{Institute for Solid State Physics, The University of Tokyo, 5-1-5 Kashiwanoha, Kashiwa, Chiba Japan 277-8581, Japan}

\begin{document}


\title{
Disorder-induced suppression of superconductivity in the Si(111)-(\SqrtBy{7}{3})-In surface: Scanning tunneling microscopy study
}


\author{Shunsuke Yoshizawa}\email{YOSHIZAWA.Shunsuke@nims.go.jp}\affiliation{\AffiliationMANA}
\author{Howon Kim}\affiliation{\AffiliationISSP}
\author{Yukio Hasegawa}\affiliation{\AffiliationISSP}
\author{Takashi Uchihashi}\affiliation{\AffiliationMANA}

\date{\today}

\begin{abstract}
The critical effect of disorder on the two-dimensional (2D) surface superconductor Si(111)-(\SqrtBy{7}{3})-In is clarified by comparing two regions with different degrees of disorder.
Low-temperature scanning tunneling microscopy measurements reveal that superconductivity is retained in the less disordered region, judging from the characteristic differential conductance ($dI/dV$) spectra and from the formation of vortices under magnetic fields.
In striking contrast, the absence of those features in the highly disordered region shows that superconductivity is strongly suppressed there. 
Analysis of observed zero-bias anomalies in $dI/dV$ spectra allows us to estimate the reduction in the transition temperature $\Tc$, which explains the fate of superconductivity in each region.
\end{abstract}

\pacs{}

\maketitle



Recently, the emergence of superconductivity has been established for several kinds of silicon surface reconstructions with metal adatoms \cite{Zhang2010,Uchihashi2011,Uchihashi2013,Yoshizawa2014a,Yamada2013}.
They are particularly interesting  because of their well-defined unique atomic structures and accessibility through standard surface science techniques such as scanning tunneling microscopy (STM) \cite{Zhang2010,Brun2014,Yoshizawa2014b}. 
One of the important features of these systems is a high sensitivity to the presence of surface defects, which originates from the atomic-scale thickness of the conducting layer.
For example, the influence of isolated surface atomic steps on supercurrents was studied by electron transport measurements \cite{Uchihashi2011} and with a low-temperature (LT) STM under magnetic fields \cite{Brun2014,Yoshizawa2014b}.
Detailed analyses in these experiments have revealed that the atomic steps can play the role of Josephson junctions.
In contrast, a high concentration of randomly distributed atomic steps and point defects should be regarded as crystalline disorder.
In this case, the disorder-induced suppression of superconductivity is expected.
Although Anderson's theory claims that superconductivity is insensitive to disorder under time-reversal symmetry \cite{Anderson1959}, electron localization inherent to the disordered two-dimensional (2D) system \cite{Abrahams1979} could lead to a strong suppression of superconductivity \cite{Ma1985,Goldman1998,Lin2015}.
Moreover, STM studies on conventional thin films of superconductor have shown that the presence of strong disorder makes the superconducting state spatially inhomogeneous \cite{Sacepe2011,Bouadim2011,Noat2013}.
However, such effects still need to be investigated for surface 2D superconductors.

In the present Rapid Communication, we successfully clarify the critical influence of  disorder on the superconductivity in the Si(111)-(\SqrtBy{7}{3})-In surface reconstruction  [referred to as (\SqrtBy{7}{3})-In] using LT STM.
Topographic STM observations show two surface regions with different degrees of disorder.
In the less disordered region, superconductivity is found to survive judging from the characteristic differential conductance ($dI/dV$) spectra and from the formation of vortices under magnetic fields.
In striking contrast, these features are absent in the highly disordered region, showing strong suppression of superconductivity.
Analysis of observed zero-bias anomalies in $dI/dV$ spectra allows us to estimate the degree of disorder and the reduction in the transition temperature $\Tc$.
The result satisfactorily explains the survival and disappearance of superconductivity found in each region.


The (\SqrtBy{7}{3})-In surface was chosen as an archetypal superconductor made of a surface reconstruction \cite{Zhang2010,Uchihashi2011,Uchihashi2013,Yamada2013,Yoshizawa2014a,Yoshizawa2014b}.
Although the existence of two (\SqrtBy{7}{3})-In phases was reported \cite{Kraft1995}, the surface atomic structure observed here is only of one kind and is identical to that in our previous studies \cite{Uchihashi2011,Uchihashi2013,Yoshizawa2014a,Yoshizawa2014b}
\footnote{The details of the atomic structure of the (\SqrtBy{7}{3})-In phase studied here are still unknown. While the ARPES data were well reproduced by band calculations based on a structure model proposed recently \cite{Park2012,Uchida2013}, the experimental STM images are different from the simulated image for the same model.}.
In the following, the physical parameters needed for analysis will be taken from the angle-resolved photoemission spectroscopy (ARPES) study on a (\SqrtBy{7}{3})-In sample prepared in the same way as ours \cite{Rotenberg2003}, with an effective mass of $m^* = 1.09 m_{\rm e}$ and a Fermi energy of $\EF = 6.9$ eV. 
These give the Fermi wave number $\kF = 1.41$ \AA$^{-1}$ and the Fermi velocity $\vF = 1.49\times10^6$ m/s.

The experiments were performed using a LT STM system.
First a Si(111) substrate was flashed several times to prepare a clean surface with the $7 \times 7$ reconstruction.
Indium was thermally deposited onto the surface at room temperature and annealed to obtain the \SqrtBy{7}{3} phase.
Here we employed an annealing condition of ${\sim}400$ ${}^\circ$C for 5 min, which was lower than that of ${\sim}570$ ${}^\circ$C for 3 s used in the previous study \cite{Yoshizawa2014b}.
This decrease in annealing temperature resulted in a substantial increase in the density of surface defects. 
The presence of the \SqrtBy{7}{3} phase was confirmed by reflection high-energy electron diffraction (RHEED) patterns.
The sample was then transferred to the LT STM stage where it was cooled down to ${\sim}0.5$ K, which is sufficiently lower than $\Tc \sim 3$ K \cite{Zhang2010,Uchihashi2011,Uchihashi2013,Yamada2013,Yoshizawa2014a}.
Differential conductance $dI/dV$ ($I$: tunneling current; $V$: sample bias voltage) was measured by using a lock-in amplifier with a small ac bias modulation.
Zero-bias conductance (ZBC), i.e., $dI/dV$ at $V = 0$, was measured while the feedback was off after the tip-sample separation was fixed.
Magnetic fields were applied in the direction perpendicular to the sample surface using a superconducting solenoid magnet.


\begin{figure}
\begin{center}
\includegraphics[width=\columnwidth]{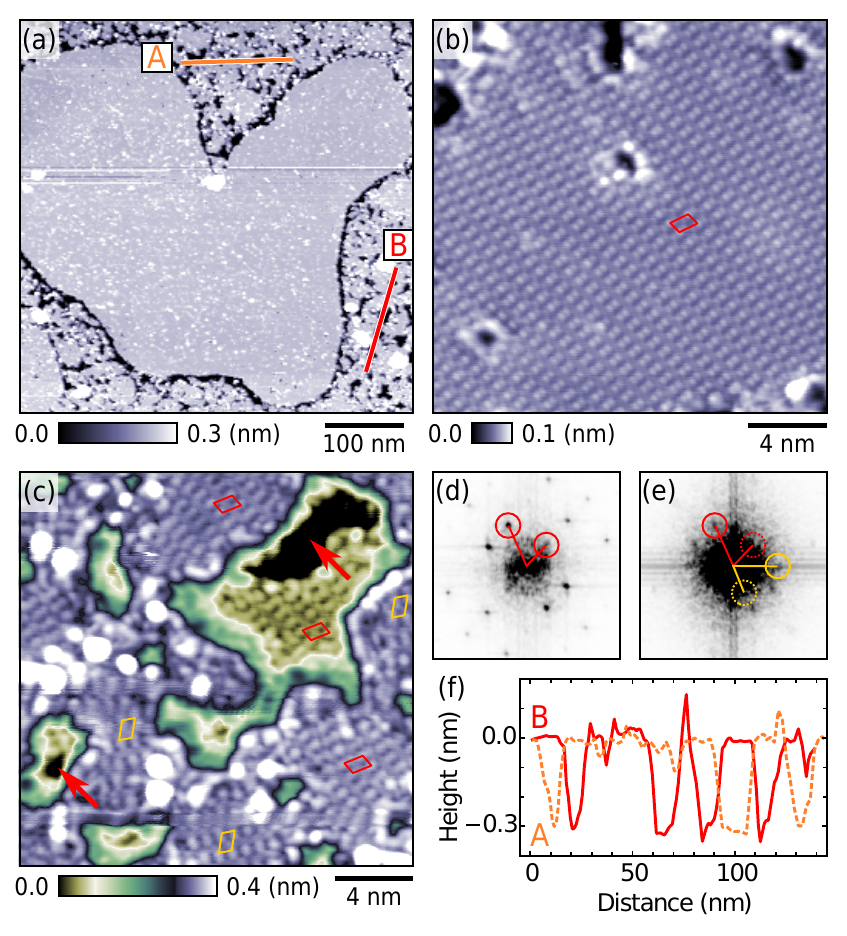}
\end{center}
\caption{\label{fig_1} (Color online)
(a) STM image of a 500 nm $\times$ 500 nm region on the surface (setpoint: 10 pA at $+50$ mV).
(b) STM images of a 20 nm $\times$ 20 nm region on a flat region 
(setpoint: 50 pA at $+500$ mV).
Note that the void defects in this image appear as white dots in (a) and in Fig. \ref{fig_2}(b) because of insufficient spatial resolutions in the latter images.
(c) STM images of a 20 nm $\times$ 20 nm region on a rough region (setpoint: 50 pA at $+500$ mV).
The arrows indicate the holes where In films did not grow.
The parallelograms in (b) and (c) represent \SqrtBy{7}{3} unit cells. 
(d), (e) Fourier transforms of the STM images (b) and (c), respectively. 
The circles indicate spots corresponding to the \SqrtBy{7}{3} periodicity.
(f) Line profiles measured along the solid lines in the STM image (a).
}
\end{figure}

First, we characterize the morphology of the sample surface.
Figure \ref{fig_1}(a) shows a representative STM image in which the flat central area and the surrounding defective region are visible.
In the following, they are referred to as ``flat'' and ``rough'' regions, respectively.
An atomic-resolution image of the flat region reveals the presence of a well-ordered \SqrtBy{7}{3} reconstruction, the unit cell of which is indicated as the parallelogram [Fig. \ref{fig_1}(b)].
The image also includes a low density of small void defects; the average spacing between the voids was found to be ${\sim}7$ nm from a larger STM image.
Accordingly, the Fourier transform (FT) of the image exhibits clear spots corresponding to the \SqrtBy{7}{3} periodicity [Fig. \ref{fig_1}(d)].
In contrast, the rough region is characterized by a high density of defects as evidenced by an atomic-resolution image with the same scale [Fig. \ref{fig_1}(c)].
Nevertheless, the \SqrtBy{7}{3} reconstruction is locally preserved within small domains 5--10 nm in size (see the parallelograms).
The presence of the \SqrtBy{7}{3} periodicity is also confirmed by its FT pattern, in which the corresponding weak spots are visible [Fig. \ref{fig_1}(e)]. 
The line profiles taken across the domains reveal that they differ in height by 0.3 nm [Fig. \ref{fig_1}(f)], indicating that the domains are separated by atomic steps of silicon.
There are often holes along the step edges, as shown by the arrows in Fig. \ref{fig_1}(c). Such structures are commonly observed along the atomic steps of the (\SqrtBy{7}{3})-In surface \cite{Yoshizawa2014b}.
To summarize, both flat and rough regions are \SqrtBy{7}{3} surfaces, but with very different degrees of crystalline disorder.
This atomic-scale characterization of disorder became possible because the (\SqrtBy{7}{3})-In superconductor has a clean surface that is suitable for STM.
As shown in the following, the observed morphological differences have a profound effect on superconductivity.

\begin{figure}
\begin{center}
\includegraphics[width=\columnwidth]{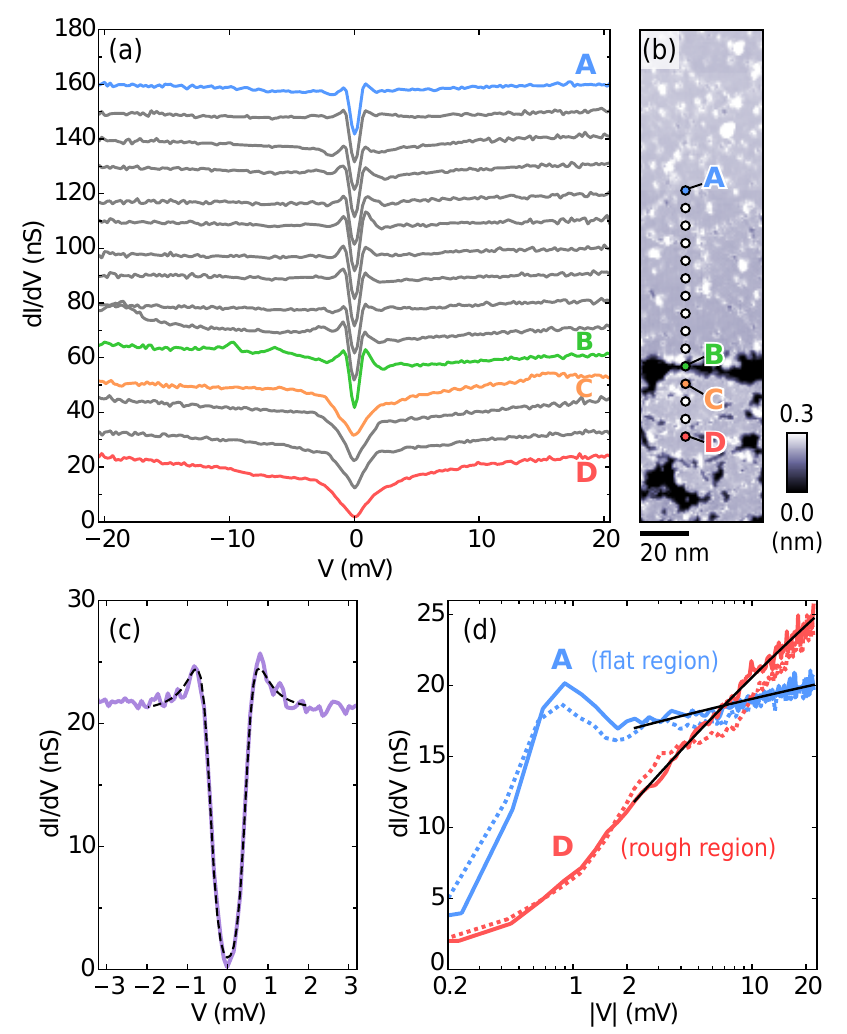}
\end{center}
\caption{\label{fig_2} (Color online)
(a) Series of tunneling spectra measured in a zero magnetic field at 15 points
indicated in the STM image in (b).
The curves are shifted from each other by 4 nS for clarity
(setpoint: 400 pA at $+20$ mV; bias modulation: 200 \uVrms{} at 610 Hz).
(b) STM image of a 50 nm $\times$ 200 nm region including a boundary between a flat region and a rough region (setpoint: $10$ pA at $+50$ mV).
The white dots in the flat region correspond to the void defects in Fig. \ref{fig_1}(b).
(c) Representative tunneling spectrum of the flat region. 
The dashed curve is the result of fitting by the Dynes formula 
(setpoint: $400$ pA at $+20$ mV; bias modulation: 50 \uVrms{} at 610 Hz,).
(d) Semilogarithmic plot of the tunneling spectra at points $A$ and $D$.
The solid (dotted) curves represent the data on the positive (negative) biases.
The black solid lines are the results of fitting by Eq. (\ref{eq:rho}).
}
\end{figure}

Figure \ref{fig_2}(a) depicts a series of $dI/dV$ spectra taken across a boundary between the flat and rough regions.
The spectral locations are indicated in the STM image of Fig. \ref{fig_2}(b).
The $dI/dV$ spectra taken in the flat region ($A$-$B$) are characteristic of the superconducting state; each spectrum exhibits an  energy gap at $V = 0$ and two coherence peaks at the gap edges.
Despite the presence of void defects, the magnitude of the energy gap was found to be almost constant within the flat region.
Figure \ref{fig_2}(c) shows a representative spectrum in the flat region measured with a higher energy resolution.
Fitting the spectrum to the Dynes formula \cite{Dynes1978} yields an energy gap $\Delta = 0.45$ meV, a broadening parameter $\Gamma = 0.08$ meV, and an electron temperature $T = 1.28$ K.
The obtained $\Delta$ is ${\sim}20$\% smaller than the value of 0.57 meV reported previously \cite{Zhang2010}.
The reason for this discrepancy will be discussed later.
In contrast, the $dI/dV$ spectra taken in the rough region exhibit only a much broader dip structure around $V = 0$ [see $C$-$D$ in Fig. \ref{fig_2}(a)].
The absence of the coherence peaks in these spectra indicates the disappearance of superconductivity in the rough region.


\begin{figure}
\begin{center}
\includegraphics[width=\columnwidth]{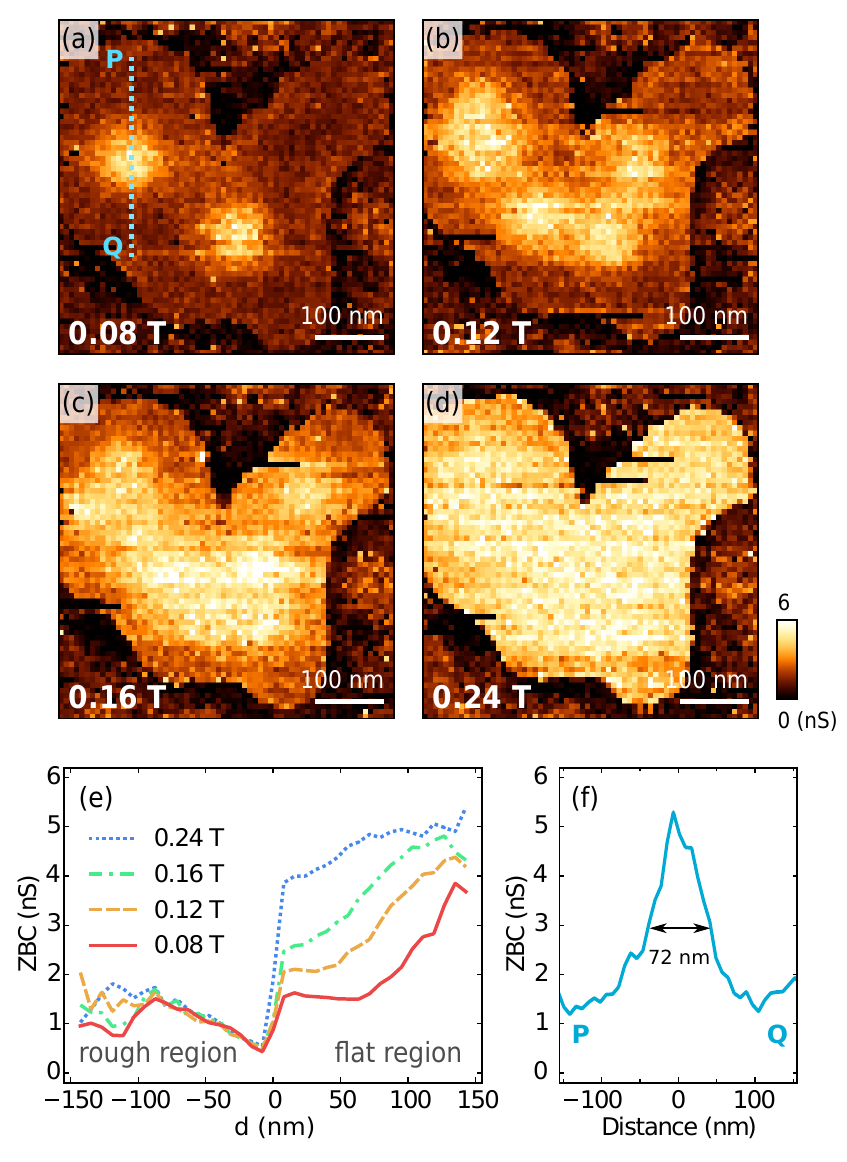}
\end{center}
\caption{\label{fig_3} (Color online)
(a)--(d) ZBC images in magnetic fields of $\Bext = 0.08$, 0.12, 0.18, and 0.24 T.
The field of view is the same as that of Fig. \ref{fig_1}(a).
Vortex cores are imaged as bright spots in the flat region
(setpoint: 200 pA at $+20$ mV; bias modulation: 200 \uVrms{} at 610 Hz).
(e) Average ZBC plotted as a function of distance $d$ from the boundary of the flat region.
To plot these curves, ZBC was averaged over all the points with a given value of $d$, where the sign of $d$ is defined to be positive (negative) for the flat (rough) region.
(f) Line profile of the ZBC image along the line $P$-$Q$ across a vortex core in (a).
}
\end{figure}

The above findings were corroborated by observing ZBC images of the same area in different magnetic fields [Figs. \ref{fig_3}(a)--\ref{fig_3}(d)].
At the lowest field of $\Bext = 0.08$ T, two bright spots corresponding to high ZBC were observed in the flat region [Fig. \ref{fig_3}(a)] and they increased in number and overlapped for $\Bext = 0.12$ and 0.16 T [Figs. \ref{fig_3}(b) and \ref{fig_3}(c)].
Thus the bright spots are attributed to vortex cores \cite{Nishio2008,Cren2009,Cren2011,Tominaga2012,Tominaga2013} and superconductivity is evidenced in the flat region.
At $\Bext = 0.24$ T, ZBC was saturated over the flat region, meaning that the superconductivity was completely suppressed [Fig. \ref{fig_3}(d)].
On the contrary, ZBC was not affected by these magnetic fields in the rough region.
This is clearly visible in Fig. \ref{fig_3}(e), where the averaged ZBC is plotted as a function of distance $d$ from the boundary between the two regions for each magnetic field.
While ZBC increases with increasing magnetic field in the flat region ($d>0$), reflecting the suppression of superconductivity in the vortex cores, ZBC is almost constant in the rough region ($d<0$).
This strongly indicates that the superconductivity was already suppressed in the rough region and that the dip structure observed for $C$-$D$ in Fig. \ref{fig_2}(a) cannot be attributed to superconductivity.

Considering the high degree of disorder in the rough region, this type of dip structure is ascribed to the zero-bias anomaly (ZBA) originating from electron-electron interactions enhanced by electron scatterings \cite{Altshuler1979,Altshuler1980,Altshuler1985}. 
In 2D systems, the ZBA is known to exhibit a logarithmic bias dependence \cite{Imry1982,White1985,Gershenzon1986,Hsu1994,Butko2000,Bielejec2001,Sherman2012,SerrierGarcia2013,Noat2013}. 
According to perturbation theory calculations  \cite{Altshuler1980,Altshuler1985}, the correction to the density of states  $\delta\rho$  to the unperturbed value $\rho = m^*/(\pi \hbar^2)$ due to this effect has the following form,
\begin{equation}
\frac{\delta\rho(\varepsilon)}{\rho} = 1 - \frac{1}{4 \pi \kF l}
\ln\left(\frac{\varepsilon}{D^2\kappa^4\hbar\tau}\right)
\ln\left(\frac{\tau \varepsilon}{\hbar}\right),\label{eq:rho}
\end{equation}
with $\varepsilon \equiv \max\{|eV|, \kB T\}$.
Here, $l$ is the elastic mean free path, $\tau \equiv l/\vF$ the elastic scattering time, $D \equiv (1/2)\vF l$ the 2D diffusion constant, and $\kappa \equiv e^2 \rho /(2\varepsilon_0)$ the 2D inverse screening length.
In the energy range of 1--20 meV,
the first logarithmic factor of the right hand side is roughly constant,
and the ZBA has a $\ln(V)$ bias dependence coming mainly from the second logarithmic factor
\footnote{This is explained as follows:
As $l \sim 1$ nm, $D^2\kappa^4\hbar\tau$ and $\hbar/\tau$ are estimated to be ${\sim}1$ MeV and ${\sim}1$ eV, respectively.
For the energy range of 1--20 meV, where the tunneling spectrum exhibited a clear $\ln(V)$ bias dependence, $\ln[\varepsilon/(D^2\kappa^4\hbar\tau)]$ varies by ${\sim}15$\%, while $\ln[\varepsilon/(\hbar/\tau)]$ by ${\sim}45$\%.
Hence, the $\ln(V)$ bias dependence is mainly from the second logarithmic factor of Eq. (1) and a possible $[\ln(V)]^2$ dependence is hardly discernible within this energy range.}.
Figure \ref{fig_2}(d) replots the $dI/dV$ spectra taken well inside the rough region (point $D$) and the flat region (point $A$). 
The spectrum at $D$ exhibits a clear $\ln(V)$ bias dependence above 1 mV, which supports our interpretation of the dip structure as a ZBA.
Assuming $\kF = 1.41$ \AA$^{-1}$ and  $\vF = 1.49\times10^6$ m/s as mentioned earlier \cite{Rotenberg2003}, $l=1.05\pm 0.12$ nm and $\kF l=15\pm 2$ are obtained by fitting Eq. (\ref{eq:rho}) to the spectrum (fitting range: 2 mV $< V <$ 20 mV).
Here, $\kF l$ is the dimensionless measure of the degree of disorder.
Thus the rough region is metallic in the sense that $\kF l \gg 1$.
This is consistent with the absence of a Coulomb gap in the spectra, which should be observed for an insulating state with $\kF l \ll 1$ \cite{Butko2000,Bielejec2001,Efros1975,Efros1976}.
We note that the $dI/dV$ taken at $A$ in the flat region also shows a $\ln(V)$ dependence above the superconducting coherence peaks, although its dependence is much weaker.
This means that a weak disorder-induced ZBA coexists with superconductivity in the flat region. 
A similar fitting analysis gives $l=2.07\pm 0.56$ nm and $\kF l=29\pm 8$, which are larger than those for $D$.
The obtained values of $l$ and $\kF l$ are summarized in  Table I, together with other parameters.

\begin{table}
\caption{Parameters determined from fitting analyses for the spectra shown in Fig. 2(a).
The errors come from the standard deviations of $l$ values determined at various positions.}
\begin{tabular}{lccccc} \hline\hline
&  $\Delta$ (meV) & $l$ (nm) & $\kF l$ & $\Tc$ (K) & $\xi$ (nm) \\ \hline
Flat region     & 0.45 & $2.07\pm0.56$ & $29\pm8$  & $1.4^{+0.4}_{-0.6}$ & $38 \pm5$ \\ 
Rough region & NA  & $1.05\pm0.12$  & $15\pm2$  & $0.19^{+0.15}_{-0.13}$ & NA \\ \hline\hline
\end{tabular}
\end{table}


The strong suppression of superconductivity in the rough region can be explained in terms of the combined effects of electron-electron interactions and disorder \cite{Maekawa1982,Takagi1982,Finkelstein1987}.
According to the theory, the decrease in $\Tc$  is given by \cite{Finkelstein1987}
\begin{equation}\begin{split}
& \frac{\Tc}{T_{\rm c0}} = \exp(-1/\gamma) \\
& \times \left[\left(1+\frac{\sqrt{t/2}}{\gamma-t/4}\right) 
\left(1 - \frac{\sqrt{t/2}}{\gamma-t/4}\right)^{-1} \right]^{1/\sqrt{2t}},\label{eq:tc}
\end{split}\end{equation}
where  $T_{\rm c0}$ is $\Tc$ in the absence of disorder, $\gamma \equiv 1/\ln(\kB T_{\rm c0}\tau/\hbar)$, and $t \equiv 1/(\pi\kF l)$.
Using $\kF l$ obtained above and $T_{\rm c0} = 3$ K, $\Tc$ for the rough region is estimated to be $0.19^{+0.15}_{-0.13}$ K.
This value is sufficiently lower than the experimental temperature of $T \sim 0.5$ K.
Thus the absence of superconductivity in the rough region is rationalized. 
In contrast, $\Tc$ for the flat region is estimated to be $1.4^{+0.4}_{-0.6}$ K from a similar argument, consistent with the presence of superconductivity in that region.
The finding is also in line with the observed suppression of the superconducting energy gap $\Delta$ by ${\sim} 20$\%, since $\Tc$ and $\Delta$ are proportional in the Bardeen-Cooper-Schrieffer (BCS) theory.
The discrepancy between the estimated reductions in $\Tc$ and $\Delta$ may reflect the limitation of the perturbation approach used in the ZBA theory \cite{Altshuler1979,Altshuler1980,Altshuler1985} and/or the deviation of the surface states from the nearly free electron model.

Lastly, we estimate the influence of disorder on the coherence length $\xi$ in the flat region.
In a clean superconductor, $\xi$ at $T= 0$ is approximately equal to the BCS value defined by $\xi_0 \equiv \hbar \vF / (\pi \Delta)$.
Here, $\xi_0=694$ nm is obtained from $\Delta=0.45$ meV in the flat region and $\vF = 1.49\times10^6$ m/s.
In the presence of disorder-induced scatterings, $\xi$ is given by $\xi \sim \sqrt{\xi_0 l}$.
Adopting $l=2.07\pm0.56$ in the flat region leads to $\xi = 38\pm5$ nm.
$\xi$ is independently determined to be 36 nm from the half width at half maximum of the ZBC profile measured over a vortex core [Fig. \ref{fig_3}(f)].
The fact that these two values of $\xi$ are nearly equal demonstrates the consistency of our analysis.


In conclusion, we have revealed that superconductivity was retained in the flat region of the \SqrtBy{7}{3} surfaces, judging from the presence of an energy gap and coherence peaks in the $dI/dV$ spectra and from the formation of vortices under magnetic fields.
In striking contrast, the absence of those features in the rough region showed that superconductivity was strongly suppressed there. 
Analysis of the zero-bias anomalies in $dI/dV$ spectra allowed us to estimate the reduction in transition temperature $\Tc$, which explained the fate of superconductivity in each region.
The present finding demonstrates the applicability of the existing theories to this class of surface 2D superconductors, and lays the groundwork for future studies on the influence of surface defects.    


This work was financially supported by JSPS under KAKENHI Grants No. 25247053, No. 25286055, and No. 26610107, and by World Premier International Research Center (WPI) Initiative on Materials Nanoarchitectonics, MEXT, Japan.


\bibliography{references}

\end{document}